\documentclass[a4paper,10pt,twoside]{cpc-hepnp}

\usepackage{multicol}
\usepackage{graphicx}
\usepackage{booktabs}
\usepackage{amssymb,bm,mathrsfs,bbm,amscd}
\usepackage[tbtags]{amsmath}
\usepackage{lastpage}
\usepackage{subfigure}

\begin{document}

\fancyhead[c]{\small Chinese Physics C~~~Vol. xx, No. x (201x) xxxxxx}
\fancyfoot[C]{\small 010201-\thepage}

\footnotetext[0]{Received 14 March 2009}

\title{Low emittance lattice for the storage ring of Turkish Light Source Facility TURKAY\thanks{Supported byTurkish Republic Ministry of Development with Grant No:DPT2006K120470. }}

\author{%
      Z. Nergiz$^{1;1)}$\email{znergiz@nigde.edu.tr}%
\quad A. Aksoy$^{2;2)}$\email{aaksoy@ankara.edu.tr}%
}
\maketitle

\address{%
$^1$ {Nigde University, Department of Physics, Faculty of Letter and Science, 51200 Nigde, Turkey}\\
$^2$ {Institute of Accelerator Technology, Ankara University, Ankara 06100, Turkey}\\
}

\begin{abstract}
TAC (Turkish Accelerator Center) project is aimed to build an accelerator center  in Turkey. The first step of the project is to construct IR-FEL facility. The second stage is to build a synchrotron radiation facility named TURKAY, which  is the third generation synchrotron radiation light source that aimed to achieve high brilliance photon beam from low emittance electron beam at 3 GeV. The electron beam parameters highly depend  on the magnetic lattice of the storage ring. In this paper a low emittance storage ring for TURKAY is proposed and beam dynamic properties of the magnetic lattice are investigated.
\end{abstract}



\footnotetext[0]{\hspace*{-3mm}\raisebox{0.3ex}{$\scriptstyle\copyright$}2013
Chinese Physical Society and the Institute of High Energy Physics
of the Chinese Academy of Sciences and the Institute
of Modern Physics of the Chinese Academy of Sciences and IOP Publishing Ltd}%

\begin{multicols}{2}

\section{Introduction}

Turkish Accelerator Center (TAC) project has been proposed to build an accelerator center in Turkey in order to use accelerators and accelerator based light sources for \underline{r}esearch \& \underline{d}evelopment (R\&D) in Turkey and its region \cite{thm}. The first step of TAC, which is under construction since 2010, is building an InfraRed Free Electron Lasers (IR-FEL) facility covering the range of 3-250 microns \cite{tarla}. The second stage is proposed to be a Synchrotron Radiation (SR) facility named TURKAY, which  is the third generation light source that aimed to achieve high brilliance photon beam from low emittance electron beam~\cite{aksoy, nergiz}.  The project is now entering its detailed design phase: after the completion of the conceptual design report.

The third generation light sources can be classified as low energy, intermediate energy (2-4 GeV) and high energy light sources. The disadvantage of low energy sources is that high quality x-ray beam can not be produced. In contrast to cost of low energy sources the high energy sources are quite expensive. Thermal loading in the beam
 line equipment is also another problem of high energy light sources. In the last decades the intermediate and high brightness light sources have become more preferred light sources. Photon energy ranged from few ev to 100 keV can be produced by the development of the insertion devices technology. Therefore, the intermediate energy light sources require relatively lower construction budget and can provide relatively wide photon energy range. According to this comparisons and the requirements of synchrotron radiation users from the light source user meetings arranged by TAC project, the main goals of TURKAY are determined. Therefore,  it has been decided to reach very low emittance value at 3 GeV electron beam energy and relatively short storage ring circumference. The emittance value of the storage ring should be below 1 nm rad.

\section{Optical structure of storage ring}

One of the important parameters of a light source is the brilliance which is defined as the number of photons emitted per second, per photon energy bandwidth, per solid angle and per unit source size. The brilliance can be expressed as~\cite{wille, bilderback}
\begin{equation}\label{brilliance}
B=\frac{dN/dt}{4\pi^{2}\sigma_{x}\sigma_{x'}\sigma_{y}\sigma_{y'}\frac{\Delta \omega}{\omega} }\thickapprox \frac{flux}{4\pi^{2} K \varepsilon_{x}^2}
\end{equation}
where  K is the coupling constant and $\varepsilon_{x}$  is the natural emittance. The flux depends on the beam current, undulator period and undulator deflection parameter.
So, the brilliance of a storage ring is strongly related to  the natural emittance and in order to maximize it, the horizontal and vertical emittances must be as small as possible.

Emittance is defined as the area of the phase ellipse and it is characteristic of a storage ring.  The emittance is correlated to the beam cross section  via
\begin{equation}\label{sigma}
\sigma_{x, y}=\sqrt{\varepsilon_{x, y} \beta_{x, y}}
\end{equation}
where $\beta_{x,y}$ is one of the Twiss parameters which is a position dependent quantity and it describes the beam focusing at that point.

In a storage ring, the optical structure is basic part of the machine and the main parameters of the stored beam is determined by the magnetic lattice. The main consideration of the magnetic lattice is to achieve low emittance electron beam in order to obtain high brilliance photon beam. 

The zero current natural emittance in an electron storage ring governed by 
\begin{equation}\label{eq:natemit}
\varepsilon_{0}\sim\frac{E^{2}}{N_{s}^{3}N_{d}^{3}}
\end{equation}
where E is the electron beam energy, $N_{s}$ is the number of sector and $N_{d}$ is the number of the bending magnets~\cite{borland}. According to this equation, it is decided to work on multi bend lattice.
 
Four bending magnet lattice structure is designed for main cell to get low emittance beam. The main cell consists of 4 bending magnets and 4 different type (16) quadrupole magnets. Figure~\ref{twiss} shows the betatron and dispersion functions of main cell.  
\begin{center}
\includegraphics[width=6cm, angle=270]{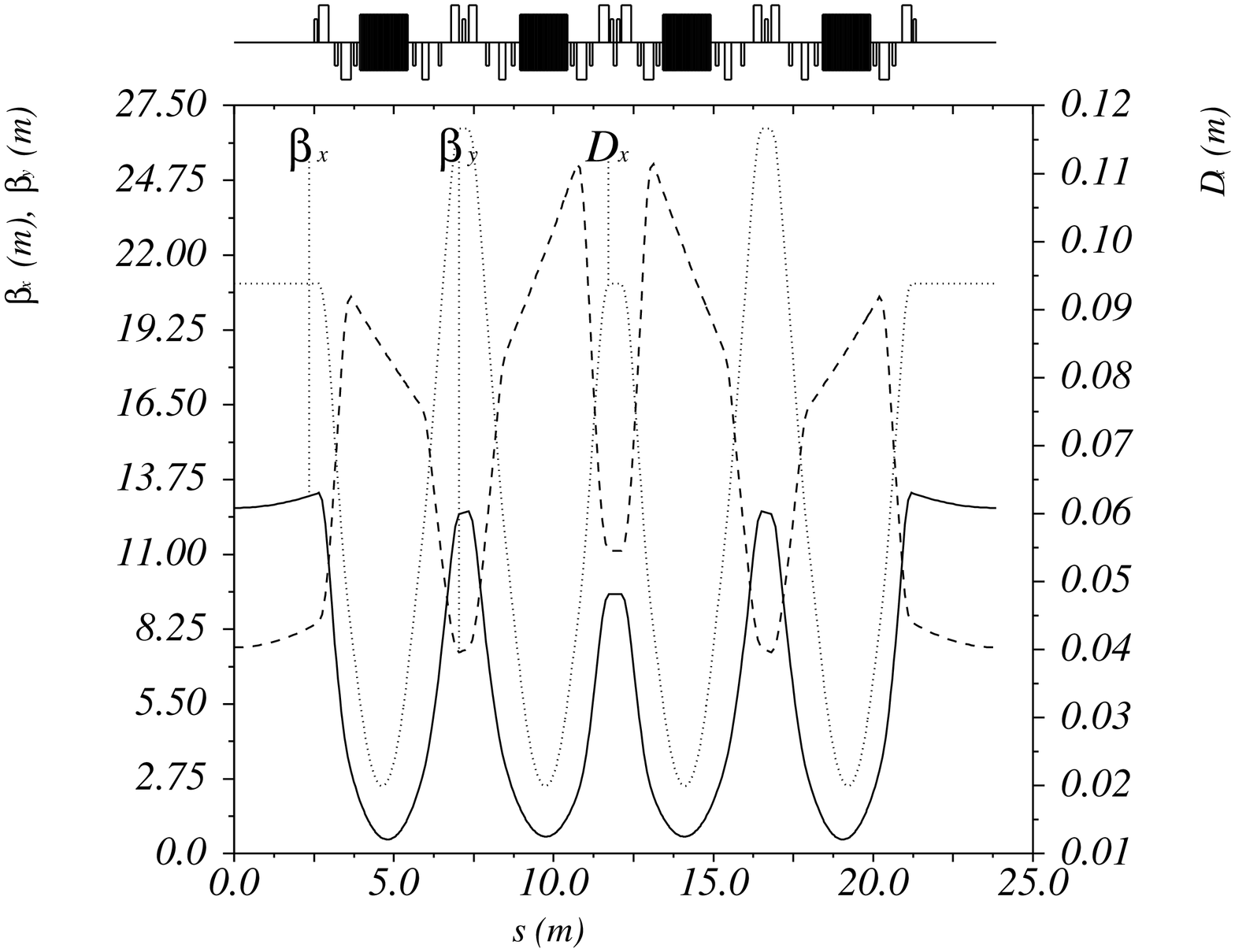}
\figcaption{\label{twiss}  Betatron and dispersion functions in main cell for finite dispersion mode.}
\end{center}
The ring has finite dispersion in straight sections and its emittance value is 0.51 nm rad. The lattice
optics can be easily tuned to achromatic lattice by adjusting the strength of the quadrupoles. In achromatic case 
 the emittance is 0.93 nm rad. 

The storage ring is composed of 20 main cells and its circumference is 477 m.  The working point is chosen where the resonance driving terms and tune shifts are as small as possible. 7 families of sextupoles are used to correct chromaticity and to compensate nonlinearities. In total, there are 80 bending magnets, 320 quadrupoles and 460 sextupoles in the whole ring.
The length of straight sections is considered to be 5 m for rf cavity, insertion devices and injection requirements. Number of the straight sections is 20; two of them for injection and rf cavity and the rest sections are for undulators. Main consideration on the number of bending magnets and sections comes from the limitation of the circumference and budget.
The bending magnets are 1.5 m long and the magnetic field in the bending magnets is 0.52 T. The advantage of the low field of the bending magnets is low rf power requirement. 

 The main parameters of storage ring are listed in Table~\ref{storageparam} for finite dispersion and achromatic modes.   OPA~\cite{opa}, MADX~\cite{madx} and BEAMOPTICS~\cite{bopt} codes are used to design magnetic lattice of storage ring.

\begin{center}
\tabcaption{ \label{storageparam}  The main parameters of storage ring.}
\footnotesize
\begin{tabular*}{80mm}{l@{\extracolsep{\fill}}ccc}
\toprule Parameters   	                & F. D. Mode	& Achr. Mode \\
\hline
    Energy  (GeV)                      & 3.0	           & 3.0\\
    Circumference (m)                          &	477               & 477	    \\
  Beam Current  (mA)                        &	500	    	& 500\\
  H. Emittance (nm rad)                 &	0.51	          & 0.93\\
V. Emittance (nm rad)                 &	0.0051	 &0.0090   \\
  Energy Loss/part./turn (keV)                   &	375.1	          & 375.1\\
Max.$ \beta_{x}$ (m)                         &	12.7	          & 15.4\\
Max. $\beta_{y}$ (m)                         &	25.3	          &28.1\\
$\beta_{x}$ in the mid. of str. sec. (m)   & 12.7	          &15.4 \\
$\beta_{y}$ in the mid. of str. sec.  (m) & 7.5	          & 9.8\\
$D_{x}$ in the mid. of str. sec. (m)   &	0.09	          & 0.00\\
Betatron Tunes $Q_{x}/Q_{y}$   &    31.19/6.15	& 36.24/6.18         \\ 
Natural Chromaticity  $\xi_{x}/\xi_{y}$    &-70/-38    &   -84/-43\\
Corrected Chromaticity   $\xi_{x}/\xi_{y}$  & 0.0/0.0	& 0.0/0.0\\
Number of straight sections   & 20	           &  20\\
Length of straight section (m)          &5	          & 5\\
RMS  Energy Spread  ($\% $)                  & 0.05	           & 0.058\\
Damping Time H/V/L  (ms)                    &26.9/26.9 /13.4 &  25.5/25.5/12.7 \\
RF Voltage  (MV)                         &3.5	            &  3.5\\
RF Frequency (MHz)                        &   500                 & 500\\
Harmonic Number      &795	             & 795\\
Bunch Charge  (nC)                        & 1	                      & 1 \\
RMS bunch length (mm)                         &2.1                     &  2.1\\
Momentum Acceptance ($\% $)              &4.8                       & 5.6 \\
Mom. Compaction Factor                  &  0.00032           & 0.00025\\
Coupling  ($\%$)                     & 1                       & 1 \\
Touschek Lifetime (h)                               &10.5                    & 18.6\\
Radiation Integrals  & &\\
I1 (m)                        & 0.135068624        & 0.1234583687            \\
I2 ($m^{-1}$)            &  0.328986813        & 0.3289868134           \\
I3 ($m^{-2}$)            &  0.017225709        & 0.0172257092            \\
I4 ($m^{-1}$)            &  0.000370301       & 0.00033846813          \\
I5  ($m^{-1}$)            &  0.000023149        & 0.0000248347      \\
\bottomrule
\end{tabular*}
\vspace{0mm}
\end{center}
\vspace{0mm}
In a low emittance lattice, strong focusing leads to high chromaticity. Therefore, it is needed to use strong sextupoles to correct the
natural chromaticity that limits the dynamic aperture. Two sextupoles are used for chromaticity corrections and those sextupoles leads to tune shift. They should be placed in suitable places to minimize the nonlinear effects. One of them ($k_{2}>0$) is located to the beginning of the lattice where $\beta_{x}$ is large and $\beta_{y}$ is small. The second one ($k_{2}<0$) is located where the point $\beta_{y}$ is large and $\beta_{x}$ is small. The purpose of the other 5 sextupoles are to compensate the nonlinear effects of the chromatic sextupoles. Their locations and strengths are determined by the use of OPA simulation code and the goals are to keep the dynamical aperture as possible as large and 
to keep the particles away from the resonance lines.  

\section{Nonlinear beam dynamic}
\label{sec:3}

The dynamic aperture is the stable transverse area for the particle in which it can execute
its oscillations safely without getting defused or lost. Storage rings require a large dynamic aperture in order to
achieve good injection efficiency and good beam lifetime. Large dynamic aperture is also an indication of high beam stability. ELEGANT code~\cite{elegant} is used to determine dynamical aperture. 
The dynamical aperture for different momentum offset and for several magnet error situations are demonstrated in Figs.~\ref{daoff} and ~\ref{damerr}, respectively. It can be seen from the figures that the dynamical aperture is large enough for such a low emittance ring and effects of the magnet error is not high. 

\begin{center}
\includegraphics[width=8cm]{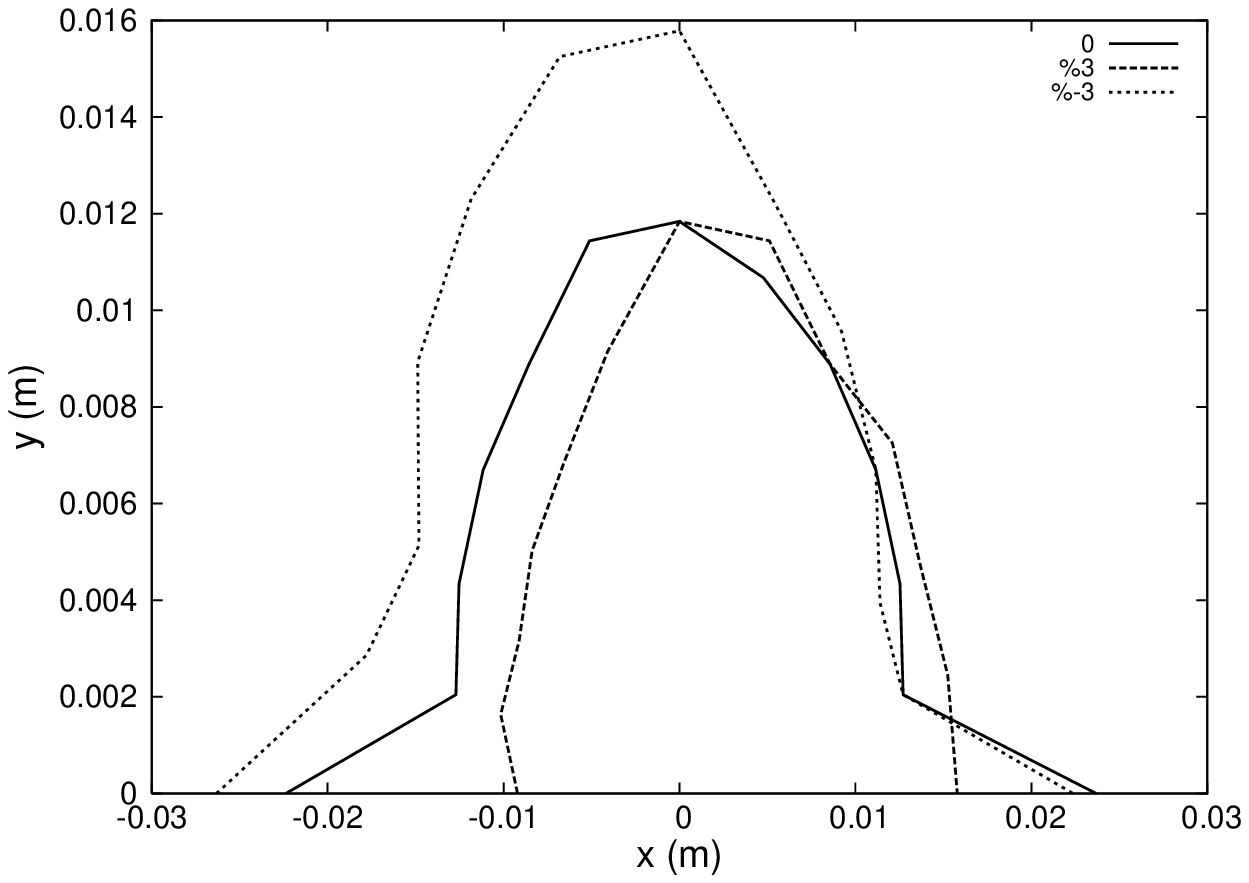}
\figcaption{\label{daoff}  Dynamical aperture of storage ring for different momentum offsets (1024 turn, magnet errors is zero).}
\end{center}

\begin{center}
\includegraphics[width=8cm]{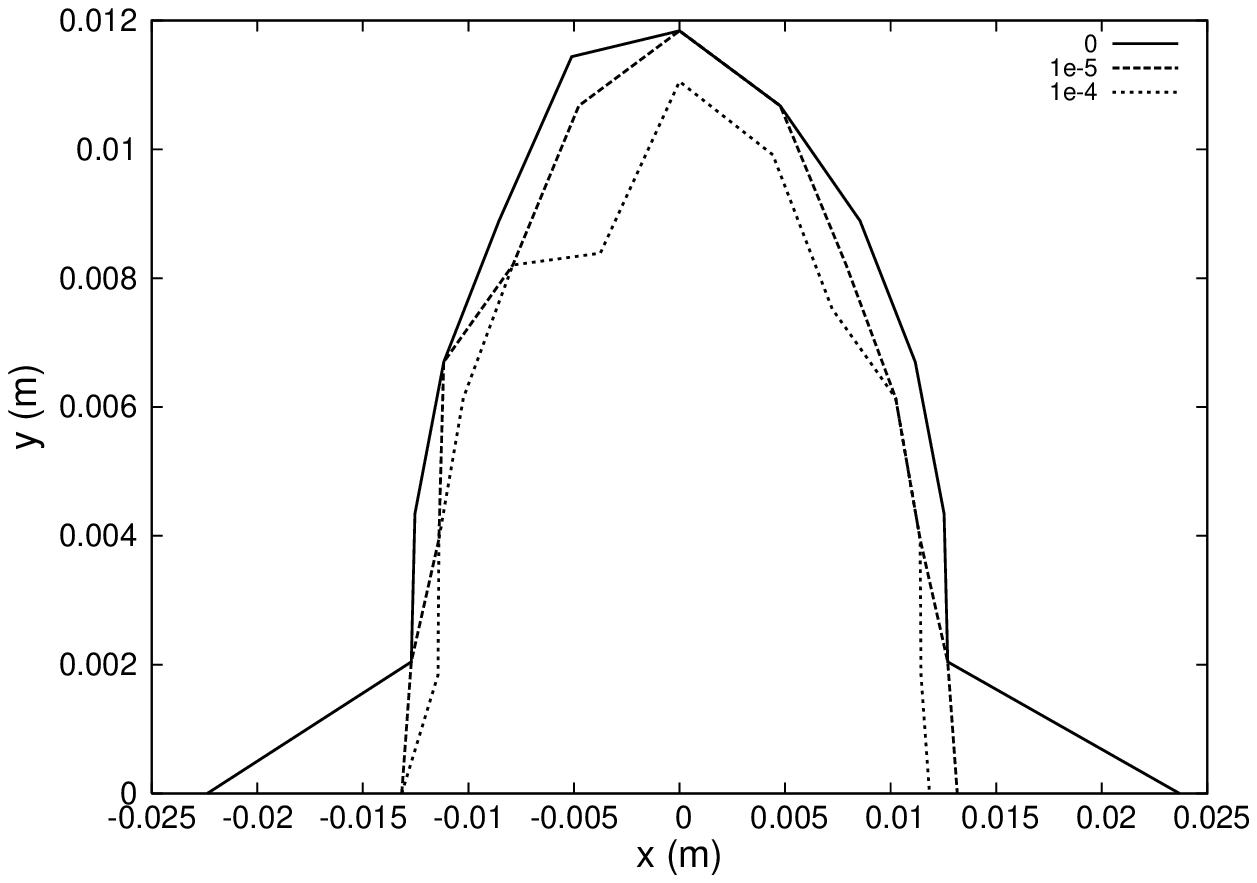}
\figcaption{\label{damerr}  Dynamical aperture of storage ring for different magnet errors (1024 turn, momentum offset is zero).}
\end{center}

Frequency Map Analysis  (FMA)~\cite{laskar} is a numerical method, which is a way to understand the effect of resonance to the dynamic aperture. For this purpose diffusion rate 
\begin{equation}\label{diffusion}
D=Log_{10}\sqrt{\Delta\nu_{x}^2+\Delta\nu_{y}^2}
\end{equation}
can be used as stability index~\cite{nadolski}. $\Delta\nu_{x}$ and  $\Delta\nu_{y}$  are transverse tune shifts of surviving particles in tracking.
 Figure~\ref{fma1} shows the dynamical aperture from FMA results of single particle tracking without synchrotron radiation. 

\begin{center}
\includegraphics[width=8cm]{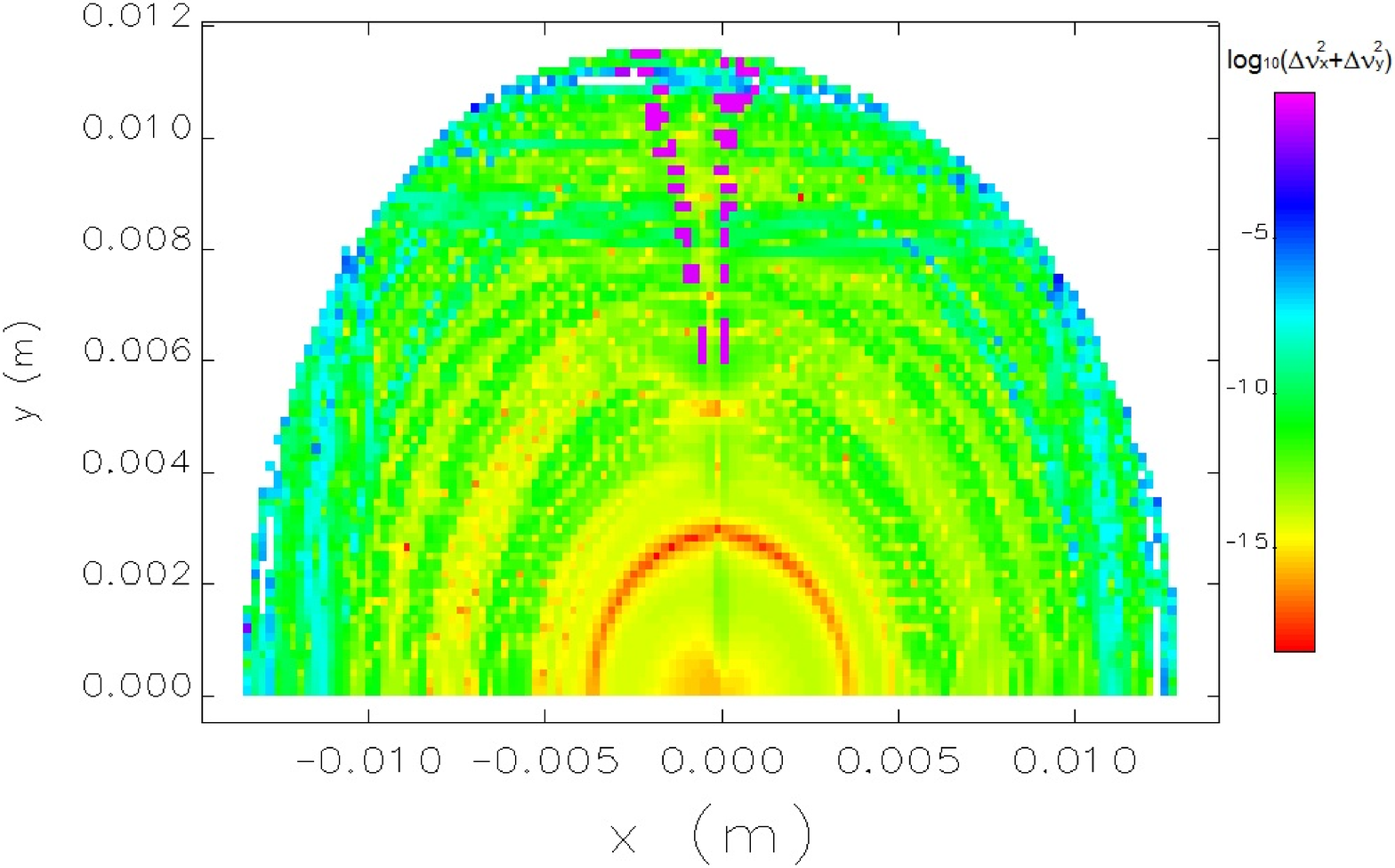}
\figcaption{\label{fma1} The effects of the resonance to dynamical aperture from Frequency Map Analyses.}
\end{center}

One of the important nonlinear issue of a low emittance storage ring is intrabeam scattering which is the coulomb scattering within a beam that usually increase the beam emittance. The horizontal beam emittance and relative energy spread with intrabeam scattering are ~\cite{bane, wolski}
\begin{equation}\label{emitibs}
\varepsilon_{x}=\frac{T_{x}}{T_{x}-\tau_{x}}\varepsilon_{x0},
\end{equation}
\begin{equation}\label{sigmaibs}
 \sigma_{\delta}=\frac{T_{x}}{T_{x}-\tau_{p}}\sigma_{\delta 0}
\end{equation}
where $T_{x}$ and $T_{p}$ are the horizontal and longitudinal intrabeam scattering growth rates, $\tau_{x}$ and $\tau_{p}$ are the horizontal and longitudinal radiation damping times and $\varepsilon_{x0}$ and $\sigma_{\delta 0}$ are the zero current horizontal emittance and relative energy spread.

The emittance value with intrabeam scattering is calculated by using ELEGANT code. The obtained emittance value is 0.70 nmrad for 500 mA average  beam current,
3.5 MV rf voltage and 500 MHz rf frequency. This emittance value is used at brilliance and flux density calculations mentioned in next section.

Another nonlinear beam dynamics effect is Touschek lifetime. It is the limitation due to Touschek scattering which describes a collision of two electrons inside a bunch with transfer of
transverse momentum into longitudinal momentum. According to the simulation results on the storage ring the Touschek lifetime is 10.5 h. It is planned to operate the machine in top-up operation mode of that aims to maintain a steady current in a storage ring by periodically injecting small amounts of current.

\section{Radiation properties}
 \label{sec:4}

The radiation spectrum is investigated by the example of some existing or planned insertion devices from other synchrotron radiation facilities~\cite{tps} with some little changes. The amended undulator parameters are sown in Table ~\ref{unduparam}.

\begin{center}
\tabcaption{ \label{unduparam}  The main parameters of insertion devices.}
\footnotesize
\begin{tabular*}{80mm}{l@{\extracolsep{\fill}}cccc}
\toprule Parameters   &  CU18 & SU15      & IU28	& U90  \\
\hline
     Period Length  (cm)                       & 1.8	& 1.5         & 2.8	  & 9  \\
    Number of Period                             & 222       &    67       & 	142      & 44 	    \\
  Min. Gap  (mm)                        & 5		& 5.6        &  7     	&  35 \\
  K$_{ymax}$                                       & 2.25	& 2.10       & 	 2.35    & 10 \\
  Length (m)                              & 4.0 	 & 1.005     &  4.0      & 4.0\\
   Photon Energy (keV)                &   1.4-20 	& 1.8-23       & 0.8-12 	 & 0.1-3  \\
  \bottomrule
\end{tabular*}
\vspace{0mm}
\end{center}
\vspace{0mm}

IU28 is in vacuum undulator with 28 mm period length. The CU18 cryogenic permanent magnet undulator  can provide brilliance value up to 
$1.7\times 10^{21}$ photon $ s^{-1} mrad^{-2} per 0.1\% BW $. SU15 the superconducting undulator with period length of 15 mm can provide photons in energy range 1.4-20 keV. U90 is conventional permanent magnet undulator with period length 9 cm. The brilliance is calculated with SPECTRA code~\cite{spectra} and obtained graphic is presented in  Fig.~\ref{brilliance}. 

\begin{center}
\includegraphics[width=8cm]{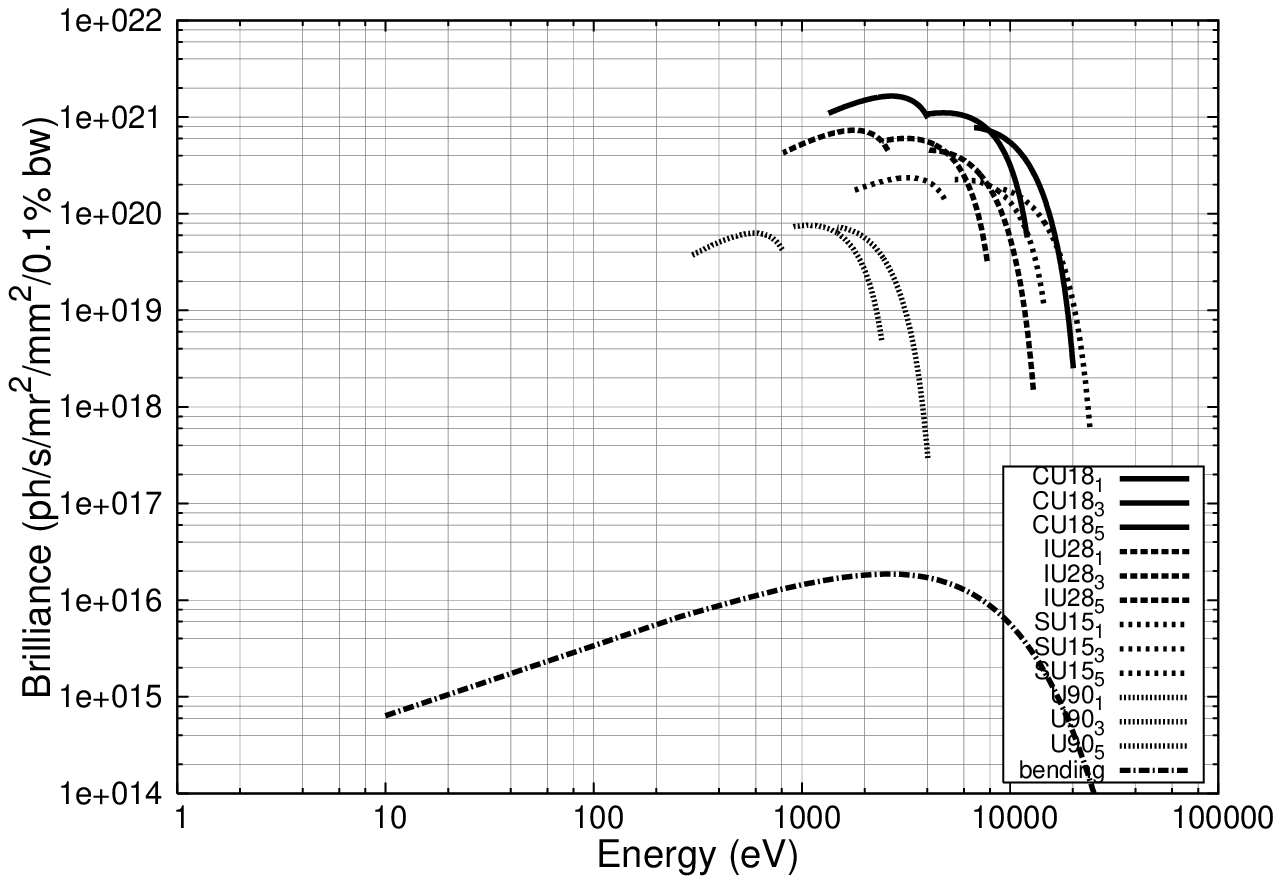}
\figcaption{\label{brilliance} Brilliance spectrum.}
\end{center}

\section {Conclusion}

The storage ring for TURKAY is designed and its storage ring parameters are presented. 0.51 nm rad emittance value is achived with relatively short circumference. 
The nonlinear effects are also investigated. It is seen that  dynamical aperture values are large enough for such a low emittance ring. The brilliance value more than $10^{21}$  $photon/s/\%0.1BW/mm^{2}/mrad^{2}$ 
can be provided by this storage ring parameters with appropriate undulators.

\section {Acknowledgment}
The authors would like to thanks to H. Wiedemann. This work was supported by Turkish Republic Ministry of Development with Grant No:DPT2006K120470.\\
\end{multicols}

\vspace{-1mm}
\centerline{\rule{80mm}{0.1pt}}
\vspace{2mm}

\begin{multicols}{2}

\end{multicols}

\clearpage

\end{document}